\documentclass[prl,aps,twocolumn,superscriptaddress,balancelastpage]{revtex4-1}

\usepackage{latexsym,amssymb,bm,bbold,amsfonts,amstext,graphicx,bbm,bm,relsize,dsfont,times}
\usepackage[dvipsnames]{xcolor}
\usepackage{enumerate}
\usepackage{amsmath}
\usepackage{amsthm}
\usepackage[unicode=true, breaklinks=false, pdfborder={0 0 1}, backref=false, colorlinks=true, linkcolor=blue, citecolor=blue]{hyperref}
\usepackage[qm]{qcircuit}
\usepackage{subfiles}

\def\BraVert{\egroup\,\mid\,\bgroup}

\usepackage{tcolorbox}

\definecolor{yellow}{HTML}{FFDB25}
\definecolor{white}{HTML}{FFFFFF}
\definecolor{light_yellow}{HTML}{FFF6B3}
\definecolor{mag}{HTML}{D33EBE}
\definecolor{light_orange}{HTML}{FFC7A5}
\definecolor{red}{HTML}{FF006C}
\definecolor{blue}{HTML}{0046FF}
\definecolor{light_green}{HTML}{A5FFD2}
\definecolor{light_blue}{HTML}{A5ECFF}
\definecolor{light_deep_blue}{HTML}{D4E2FF}
\definecolor{az}{HTML}{00B9E6}

\newcommand{\x}{\mathbf{X}}

\usepackage{tikz}

\begin{document}

\title{Stabilizing Open Quantum Batteries by Sequential Measurements}

\author{Stefano Gherardini}
\email{gherardini@lens.unifi.it}
\affiliation{\mbox{Department of Physics and Astronomy \& LENS, University of Florence,} via G. Sansone 1, I-50019 Sesto Fiorentino, Italy}
\affiliation{\mbox{INFN Sezione di Firenze}, via G. Sansone 1, I-50019 Sesto Fiorentino, Italy}

\author{Francesco Campaioli}
\affiliation{\mbox{School of Physics and Astronomy, Monash University,} Victoria 3800, Australia}

\author{Filippo Caruso}
\affiliation{\mbox{Department of Physics and Astronomy \& LENS, University of Florence,} via G. Sansone 1, I-50019 Sesto Fiorentino, Italy}

\author{Felix C. Binder}
\affiliation{\mbox{Institute for Quantum Optics and Quantum Information - IQOQI Vienna, Austrian Academy of Sciences,} Boltzmanngasse 3, 1090 Vienna, Austria}

\begin{abstract}
A quantum battery is a work reservoir that stores energy in quantum degrees of freedom. When immersed in an environment an \textit{open quantum battery} needs to be stabilized against free energy leakage into the environment. For this purpose we here propose a simple protocol that relies on projective measurement and obeys a second-law like inequality for the battery entropy production rate.
\begin{description}
\item[PACS numbers]
03.65.Yz, 05.70.Ln, 42.50.Lc
\end{description}
\end{abstract}

\maketitle

Among recent research in quantum thermodynamics~\cite{Kosloff2013,Goold2016,Vinjanampathy2016,Binder2019,BookDeffner}, the design of quantum energy storage-devices, called \emph{quantum batteries}~\cite{Campaioli2019,Alicki2013,Skrzypczyk2014,Binder2015a,Binder2016,Campaioli2017,Ferraro2017,Friis2017,Andolina2018,Farina2018,Andolina2018a,
Julia-Farre2018,Barra2019,Pintos2019ArXiv,SantosPRE2019}, is of increasing interest. So far the main focus has lied on multipartite speed-up effects in charging~\cite{Alicki2013,Binder2015a,Binder2016,Campaioli2017,Ferraro2017,Farina2018,Andolina2018a,Julia-Farre2018}, fluctuations in charging precision~\cite{Skrzypczyk2014,Friis2017,Julia-Farre2018,Pintos2019ArXiv}, and mitigating imprecise unitary control pulses~\cite{SantosPRE2019}. However, as of yet no attention has been paid to efficiently stabilizing charged quantum states, even if contributions in the area of control theory~\cite{BrifNJP2010,KochReview2016} touch upon this question both in classical~\cite{Horowitz2017} and in quantum settings~\cite{Ticozzi2013,Horowitz2015}.

In this Letter, we introduce the concept of an \emph{Open Quantum Battery} (OQB). Here, the quantum system $\mathcal{B}$, acting as a battery, interacts with the surrounding environment $\mathcal{E}$ leading to decoherence~\cite{Breuer2002}. Due to this interaction, the entropy of the battery increases~\cite{CamatiPRL2016,GherardiniEntropy,Batalhao2018} and thus unitary control pulses applied to the system are not generally sufficient to compensate such entropy production and then stabilize the system; rather, we would require a source of free-energy, such as a low-temperature heat bath. For this purpose, we propose a stabilization scheme based on a sequence of repeated quantum measurements~\cite{DePasqualePRA2017,GherardiniIQIS2018}, each of them preserving the trace of the system, i.e., no post-selection is performed, as shown in Refs.~\cite{Campisi2010PRL,Pekola2013PRL,Gherardini2018PRE}. The adoption of quantum measurements has been recently proposed for the realisation of a quantum Maxwell's demon engine~\cite{ElouardPRL2017} and in~\cite{BuffoniPRL2019} to fuel a cooling engine. Our goal is to neutralize the local increase of entropy and ensure energy-efficient control operations for implementing fast, on-demand charging/discharging and stabilization protocols, using the lowest amount of energy.

\paragraph*{Open quantum batteries.--}

A quantum battery is a finite-dimensional quantum system $\mathcal{B}$ whose energy is quantified by a bounded internal Hamiltonian $H_0$. We here consider the highest energy state $|e\rangle$ and the lowest energy state $|g\rangle$, both eigenstates of $H_0$, representing the maximally charged and discharged battery states, respectively. Hence, the battery's \emph{capacity}~\cite{Binder2016} is simply given by $E_{\textrm{max}} \equiv \text{Tr}[H_0(\rho_e-\rho_g)]$, with $\rho_e \equiv |e\rangle\!\langle e|$ and $\rho_g \equiv |g\rangle\!\langle g|$. If the battery system were perfectly isolated it would always evolve unitarily. In contrast, an \emph{open quantum battery}, when left uncontrolled, evolves under the effect of some open dynamics, i.e., $\dot{\rho}_t = -i[H_0,\rho_t]+ \mathcal{D}[\rho_t]$, where, here and below, $\hbar$ is set to $1$, $\rho_t$ denotes the density operator of the system at time $t$, and $\mathcal{D}$ is the super-operator modeling free-energy leakage due to decoherence. Equivalently, the evolution of the system can also be described by means of a time-parameterized family of completely-positive and trace-preserving (CPTP) maps $\Lambda_t:\rho_0\rightarrow\rho_t$, where $\rho_t = \Lambda_t[\rho_0]$, with steady state $\overline{\rho} \equiv \lim_{t\to\infty}\Lambda_t[\rho_0]$ and initial density operator $\rho_0$.

A charging protocol must be able to powerfully charge the battery, bringing it into the excited state $|e\rangle$ from an arbitrary initial condition, e.g.\,a thermal state at inverse temperature $\beta$ or a state in the neighborhood of $|g\rangle$, so as to maximize its \emph{ergotropy}~\cite{Allahverdyan2004}, i.e., the amount of energy that can then be unitarily extracted. While a closed quantum battery can be charged by means of cyclic unitary operations, an open quantum battery experiences non-equilibrium free-energy leakage. Hence, a unitary process no further suffices to restore the battery state or to avoid the energy losses during its dynamics. Moreover, the target state $\rho_e$ does not generally belong to the unitary controllability space~\cite{KochReview2016} of the system, especially if the charging time $t_c$ is comparable with $1/\gamma$, where $\gamma$ denotes the relevant decoherence coefficient of the super-operator $\mathcal{D}$. So, how can an open quantum battery be charged and stabilized?

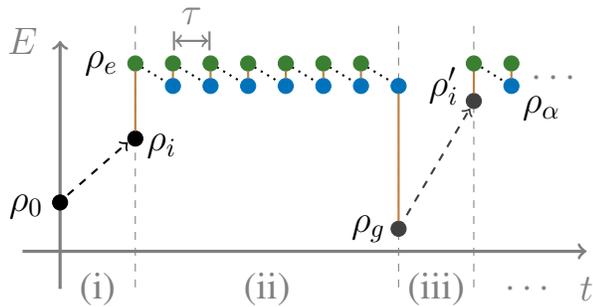
\begin{figure}[t]
\begin{tikzpicture}
 \draw[gray,very thick,->] (2.5,0) -- (10,0);
 \draw[gray,very thick,->] (3.0,-0.5) -- (3.0,2.8);
 \draw[thick,dashed,->] (3.1,0.75) -- (3.9,1.45);
 \draw[darkgray,thick,dashed,->] (7.5,0.3) -- (8.45,1.9);

\draw[thick,gray,|<->|] (4.5,2.8) -- node[above=0.1 cm] {\Large $\tau$}  (5,2.8);

%  \draw[gray,dashed] (0.5,3) -- (0.5,-1);
 \draw[gray,dashed] (4,3) -- (4,-0.5);
 \draw[gray,dashed] (7.5,3) -- (7.5,-0.5);
 \draw[gray,dashed] (8.5,3) -- (8.5,-0.5);

 \draw[black,fill=black] (3.0,0.65) circle[radius=0.1 cm];
 \draw[darkgray,fill=darkgray] (8.5,2) circle[radius=0.1 cm];
 \draw[black,fill=black] (4,1.5) circle[radius=0.1 cm];
%  \draw[OliveGreen,fill=OliveGreen] (4,2.5) circle[radius=0.1 cm];
 \draw[darkgray,fill=darkgray] (7.5,0.3) circle[radius=0.1 cm];

 \foreach \x in {0,0.5,...,3} {
 \draw[thick,dotted] (\x+4,2.5) -- (\x+4.5,2.2);
 \draw[RoyalBlue,fill=RoyalBlue] (4.5+\x,2.2) circle[radius=0.1 cm];
 \draw[OliveGreen,fill=OliveGreen] (4+\x,2.5) circle[radius=0.1 cm];
 \draw[thick,brown] (\x+4,2.3) -- (\x+4,2.4);
 };
 \draw[thick,brown] (7.5,2.1) -- (7.5,0.4);
 \draw[thick,brown] (4,2.3) -- (4,1.6);
 \draw[thick,brown] (8.5,2.4) -- (8.5,2.1);

 \draw[OliveGreen,fill=OliveGreen] (8.5,2.5) circle[radius=0.1 cm];
 \draw[OliveGreen,fill=OliveGreen] (9,2.5) circle[radius=0.1 cm];
 \draw[thick,dotted] (8.6,2.45) -- (9,2.2);
 \draw[RoyalBlue,fill=RoyalBlue] (9,2.2) circle[radius=0.1 cm];
 \draw[thick,brown] (9,2.3) -- (9,2.4);

 \draw (2.55,0.6) node {\Large $\rho_0$};
 \draw (8.1,2.2) node {\Large $\rho_i'$};
 \draw (4.35,1.4) node {\Large $\rho_i$};
 \draw (3.55,2.5) node {\Large $\rho_e$};
 \draw (7.1,0.3) node {\Large $\rho_g$};
 \draw (9.4,1.9) node {\Large $\rho_\alpha$};

 \draw[gray] (10,-0.5) node {\Large $t$};
 \draw[gray] (2.5,2.8) node {\Large $E$};

 \draw[gray] (3.5,-0.5) node {\Large (i)};
 \draw[gray] (5.75,-0.5) node {\Large (ii)};
 \draw[gray] (8,-0.5) node {\Large (iii)};
 \draw[gray] (9.25,-0.5) node {\Large $\cdots$};
 \draw[gray] (9.6,2.3) node {\Large $\cdots$};

\end{tikzpicture}
  \caption{Stabilization scheme -- single-run illustration. After the initialization step (with initial state $\rho_0$), the battery stabilization protocol consists of intermittent free evolutions and fast unitary controlled dynamics (dotted points, corresponding to steps (i) and (iii) of the procedure) and projective measurements (solid brown line, step (ii)) in time intervals of duration $\tau$. In particular, the green dots denote the maximum energy state $\rho_e$, while the blue dots represent the state $\rho_{\alpha}$ of the battery immediately before a projective measurement in the Zeno regime. $\rho_i$ and $\rho_i'$ are the nearest states to $\rho_e$ on the unitary orbit of $\rho_0$ and $\rho_g$, respectively.}
  \label{fig:scheme}
\end{figure}

\paragraph*{Stabilization scheme.--}

In this Letter, we propose a \emph{non-unitary} (NU) control protocol that counteracts the increase of entropy induced by the interaction with the environment. The control scheme is realized by a sequence of projective measurements and intermittent driving as illustrated in Fig.\,\ref{fig:scheme}. The first objective is to \textit{charge} the battery by bringing it towards the excited state $\rho_e$. The second one is to maintain the system in the neighborhood of $\rho_e$ during the time interval $[0,t_{\rm fin}]$. Now, we introduce each step of the stabilization protocol.

\textbf{(i) Initialization}: Given an input state $\rho_0$, the battery $\mathcal{B}$ is driven to that out-of-equilibrium state $\rho_i$ on its control orbit which lies closest to $\rho_e$ (e.g.\,in terms of the trace distance $T(\rho,\sigma) \equiv \frac{1}{2}\text{Tr}[\sqrt{(\rho - \sigma)^2}]$). In general, this operation combines the uncontrollable open system dynamics with Hamiltonian control $H(t)=H_0+V(t)$. While magnitude constraints on the driving may impose bounds on the maximum achievable power~\cite{Campaioli2017}, we here assume that driving fields of arbitrary magnitude are permitted. This allows for this step of the protocol to be performed sufficiently \emph{fast} so as to be considered unitary, despite the presence of decoherence. However, also regimes of slow control could be equally treated by properly adaption of existing control methods to the present scenario~\cite{Suri2018}.

\textbf{(ii) Quantum measurements and Zeno protection}: After driving $\mathcal{B}$ into the state $\rho_i$, a projective energy measurement (in the eigenbasis of $H_0$) is performed on the battery: with probability $P_{e} \equiv  \text{Tr}\left[\rho_{i}\rho_{e}\right]$ the state of $\mathcal{B}$ collapses into the excited state, while with probability $P_{g} \equiv 1 - P_{e}$ the collapse occurs into one of the other energy eigenstates. After the measurement, if $\mathcal{B}$ has collapsed into the maximum energy state $\rho_e$, then a \emph{Zeno protection protocol} is applied. The latter consists of a sequence of frequent projective measurements (again in the energy eigenbasis) at discrete periodic times with the aim of freezing the dynamics of the battery and thus stabilizing it in the excited state. As proved in~\cite{SmerziPRL2012,SchaferNATCOMM2014,MuellerPRA2016}, the time interval $\tau$ between two consecutive measurements has to be chosen according to the relation $\Delta^{2}H_{\textrm{Zeno}}\,\tau^2 \ll 1$, where $\Delta^{2}H_{\textrm{Zeno}}$ is the variance of the effective Zeno Hamiltonian $H_{\textrm{Zeno}} \equiv \rho_{e}H_{0}\rho_{e} = E_{e}\rho_{e}$ ($E_{e} \equiv {\rm Tr}[H_{0}\rho_{e}]$) with respect to the freezing state. This physically means that the battery is repeatedly brought back to the maximum energy state $\rho_e$ with a probability almost equal to one as long as $\rho_e$ -- the state to be stabilized -- and the quantum state after the evolution are statistically indistinguishable, i.e., their difference is non-detectable by any measurement device~\cite{WoottersPRD1981}. Thus, for an experimental realization of the protocol, $\tau$ needs to be significantly smaller than both the time scale of the system dynamics and the characteristic decoherence time.

\textbf{(iii) Re-initialization}: If the projective energy measurement results in one of the lower energy eigenstates, the stabilization procedure is repeated from the beginning, and unitary driving is applied as in step (i). This means that the whole procedure is repeated until the Zeno protection protocol starts.

To summarize, $\rho_e$ is an \emph{unstable} state of the battery due to interactions with the environment. Hence, in order to stabilize $\mathcal{B}$, we apply the non-unitary process given by a sequence of projective measurements. Despite the probabilistic nature of this scheme, a very high fidelity $\mathcal{F} \equiv (\,{\rm Tr}\sqrt{\sqrt{\rho_e}\rho_t\sqrt{\rho_e}}\,)^2$ in stabilizing the battery can be achieved, as shown in Fig.\,\ref{fig1_main} for an example process. Further details on numerical simulations are provided in the supplemental material (SM).

\begin{figure*}
  \includegraphics[scale=0.62]{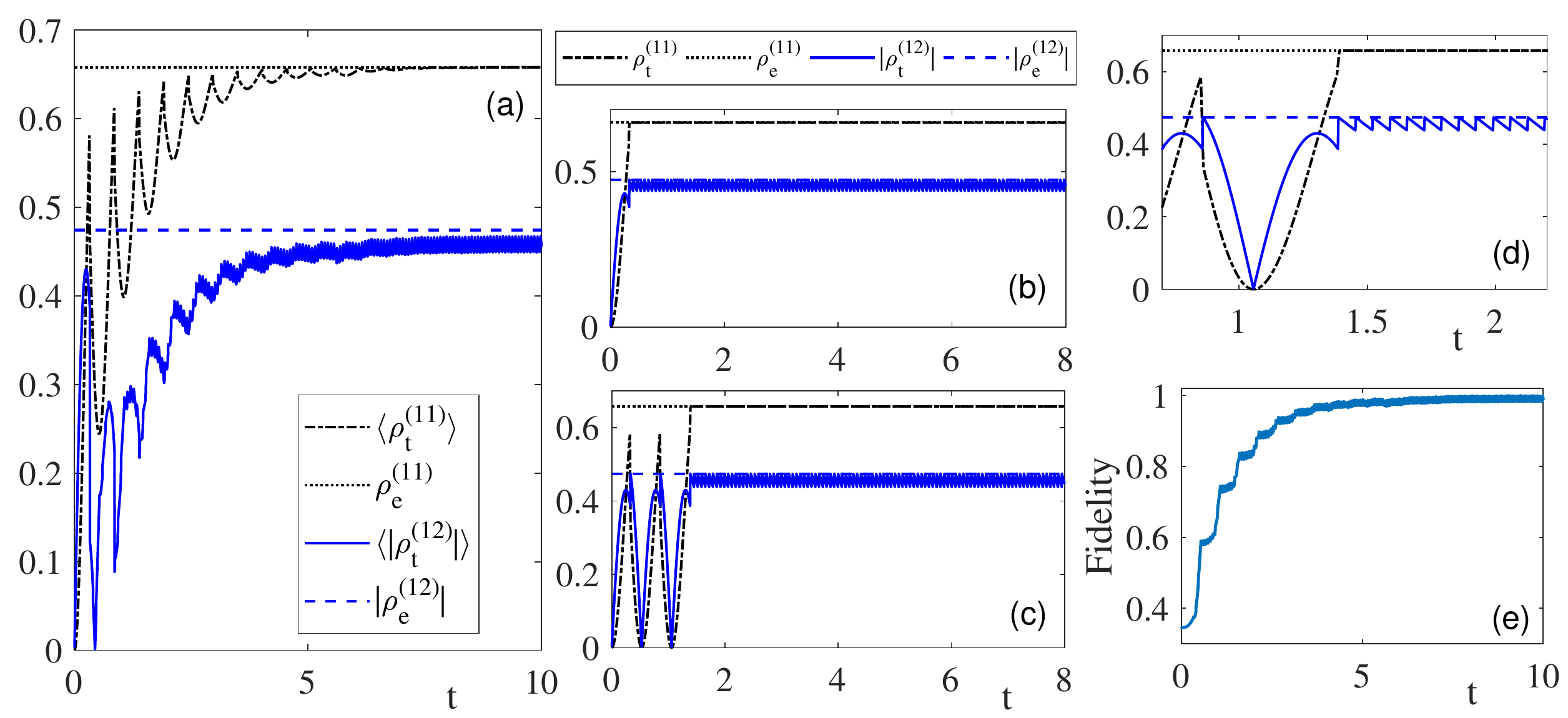}
  \caption{Stabilization scheme -- numerical results for a qubit with internal Hamiltonian $H_0 = 3\sigma_x + \sigma_z$ (in natural units). (a) Average behaviour over time of the battery density matrix, obtained by repeating the stabilization procedure $1000$ times. (b)-(c) Behaviour over time of the battery density matrix in single realizations of the scheme: being probabilistic, the charging process could require the application of more than one projective measurement. In the subplots, $(\rho_{e}^{(11)},\rho_{e}^{(12)})$ and $(\rho_{t}^{(11)},\rho_{t}^{(12)})$ are the top diagonal elements and the coherence terms, respectively, of the maximum energy state $\rho_{e}$ of the qubit-battery and of the corresponding time-evolved density matrix $\rho_{t}$. (d) Zoom of subplot\,(c) in the time interval $[0.75,2.15]$, showing the occurrence of a \emph{failure} collapse and the resulting re-initialization procedure. Further details can be found in the SM. (e) Stabilization fidelity $\mathcal{F}$ over $1000$ realization of the stabilization procedure. At $t=0$ $\mathcal{F}$ starts from a value in the range $[0.3,0.4]$ since also the initialization step has been taken into account.}
  \label{fig1_main}
\end{figure*}

\paragraph*{Performance measures.--}

For each time $t$ we define two figures of merit for the stabilization scheme. First, the ratio $\varsigma_{\textrm{stab}}(t) \equiv \langle W_{\textrm{stab}}(t)\rangle/E_{\textrm{max}}$ is named the \emph{relative stabilization cost}, where $\langle\cdot\rangle$ denotes the average over a sufficiently large number of protocol realizations and $W_{\textrm{stab}}(t)$ is the energy expended to stabilize $\mathcal{B}$. Second, we want to identify the excess cost of the stabilization procedure besides the energy cost spent to just compensate decoherence. To this end, we introduce the \emph{relative excess stabilization cost}, i.e.
\begin{equation}\label{definition_xi}
\xi_{\textrm{stab}}(t) \equiv \frac{\left|\langle W_{\textrm{stab}}(t)\rangle - \langle\Delta L(t)\rangle\right|}{E_{\textrm{max}}},
\end{equation}
where $\langle\Delta L(t)\rangle$ is the average energy leakage that would spontaneously occur if the battery were left uncontrolled. Note that $\varsigma_{\textrm{stab}}(t)$ is a cumulative energy term: it is zero when no control is applied, but can also diverge since $\langle W_{\textrm{stab}}(t)\rangle$ is an unbounded quantity. Thus, it can be easily adapted for the definition of the \emph{relative stabilization rate} $R_{\textrm{stab}} \equiv \lim_{t\to\infty}\varsigma_{\textrm{stab}}(t)/t$ which would be the same if defined in terms of $\xi_{\textrm{stab}}(t)$ rather than $\varsigma_{\textrm{stab}}(t)$, due to the long term-limit. This leads us to just consider the stabilisation power $\mathfrak{P}_{\textrm{stab}} = \dot{W}_{\textrm{stab}}$ as a performance measure in the following section.

\paragraph*{Minimum control power.--}

In this paragraph we prove a bound providing the minimum power required to stabilize the OQB. It originates from a second-law-like inequality for the battery's irreversible entropy production rate $\Sigma(\rho_t)$ (see SM for the proof). A similar result can be found in~\cite{Horowitz2015}, but concerning the energy cost to coherently control a mesoscopic quantum system. In our open-loop control framework the entropy production rate $\Sigma(\rho_t)$ equals the sum of two contributions: $\Sigma_{D}(\rho_t)$ and $\Sigma_{NU}(\rho_t)$, denoting respectively the entropy production rates due to environmental decoherence and the effect of the observer/experimenter, responsible for the non-unitary control of the battery. In particular, as discussed in the SM, the entropic contribution $\Sigma_{NU}(\rho_t)$ is equal to the time-derivative of the Shannon entropy $H(P) \equiv - \sum_{k\in\{e,g\}}P_{k}\log P_{k}$, with $P_k$'s probabilities that the battery collapses in one of the energy eigenstates. Indeed, to each of those probabilities is associated the information content of the measurement outcomes, which are stored in a classical memory~\cite{Strasberg2018}. This means that, while the measurement procedure locally reduces the battery's entropy, the reading and storing of the measurement outcomes entails an additional entropy production which cannot be neglected. According to Landauer's principle~\cite{LorenzoPRL2015,MancinoNPJ2018,Abdelkhalek2016}, the irreversible erasure of such information leads to an energy consumption, proportional to the temperature of the thermal bath used in the erasure procedure.

Once again, it is worth noting that, since $\mathcal{B}$ is affected by decoherence, the evolution of the battery admits at least one fixed point denoted as $\overline{\rho}_{\textrm{dec}}$. For the case of a qubit the steady-state $\overline{\rho}_{\textrm{dec}}$ (with no coherence in the energy eigenbasis) can always be described by an effective temperature $T_{\overline{\rho}_{\textrm{dec}}}$. The latter is interpreted as the physical temperature of a fictitious quantum system that would lead to the same decoherence effects. In particular, as shown in the SM, by defining $E(\rho_t)$ and $E_{D}(\rho_t)$, respectively, as the battery's total energy and the energy driven into $\mathcal{B}$ by the environment, the control power $\dot{W}_{\textrm{stab}}$ obeys the following inequality:
\begin{equation}\label{lower-bound_W}
\mathfrak{P}_{\textrm{stab}} \equiv \dot{W}_{\textrm{stab}}(\rho_t) \geq \dot{E}(\rho_t) - T_{\overline{\rho}_{\textrm{dec}}}\dot{S}_{D}(\rho_t),
\end{equation}
with $S_{D}$ denoting the von-Neumann entropy of the uncontrolled battery (note that $\dot{S}_{D} \propto \Sigma_{D}$). The lower-bound (\ref{lower-bound_W}) can be recast into the inequality $\dot{F}(\rho_t)\leq 0$, where $F \equiv E_{D} - T_{\overline{\rho}_{\textrm{dec}}}S_{D}$ is the battery free-energy. This inequality represents the second law of thermodynamics: the free-energy of the uncontrolled battery reduces due to the increase of the von-Neumann entropy $S_{D}$ resulting from the open systems dynamics. Therefore, the minimum value of $\dot{W}_{\textrm{stab}}$ implies the equality $\dot{F}(\rho_t)=0$, with the result that the lowest energy $W_{\textrm{stab}}^{(\textrm{min})}$ required to control $\mathcal{B}$ (note that, apart from a constant term, $W_{\textrm{stab}}^{(\textrm{min})}$ is equal to $E - T_{\overline{\rho}_{\textrm{dec}}}S_{D}$) is such that the free-energy is constant, i.e., the increase in entropy due to the environment is compensated by the control operation. Note that Eq.\,(\ref{lower-bound_W}) is valid whatever is the control action applied on $\mathcal{B}$, for this reason the symbol $\langle\cdot\rangle$ has not been used. However, for the probabilistic stabilization procedure we are proposing, the results from Eq.\,(\ref{lower-bound_W}) just hold true only on average.

\paragraph*{Energetic efficiency.--}

Returning to Eq.\,(\ref{definition_xi}), we now derive the average control energy $\langle W_{\textrm{stab}}(t)\rangle$ and environmental losses $\langle\Delta L(t)\rangle$. The battery's energy is determined by the time-independent Hamiltonian $H_0$, thus the cost $\Delta E_{\textrm{evol}}$ for the initialization of the battery and its dynamical evolution is exactly equal to ${\rm Tr}[H_0(\rho_{i}-\rho_0)]$. Indeed, for fast control (i.e., $V(t)=0$ almost $\forall t$), the integral $\int{\rm Tr}[V(t)(\rho_t - \rho_0)]dt$ is negligible and $\int\rho_{t}~dt \approx \rho_i$. On the other hand, the cost $\Delta E_{\textrm{meas}}$ of each projective measurement is given by the difference between the battery energies, respectively, after and before the measurement: with probability $P_{e}$, $\Delta E_{\textrm{meas}} = {\rm Tr}[H_0(\rho_e - \rho_{i})]$, and with probability $P_{g}$, $\Delta E_{\textrm{meas}} = {\rm Tr}[H_0(\rho_{g}-\rho_{i})]$. On average, however, there is no energetic cost associated to the measurement, i.e., $\langle\Delta E_{\textrm{meas}}\rangle = 0$, independently of $\rho_{i}$. Only the entropic cost for the erasure of the measurement information has to be considered. The same holds true for the Zeno protection protocol, whereby on average the energy cost equals $\langle\Delta E_{\textrm{Zeno}}\rangle = \overline{m}\,\beta^{-1}H(P(\rho_{\alpha}))$. Here, $\rho_{\alpha}$ is the average state just before a projection in the Zeno regime (Zeno measurement) and depends on the measurement frequency $1/\tau$, $\overline{m}$ is the average number of Zeno measurements~\footnote{For a fixed value both of $\tau$ and $t_{\rm fin}$ also the number of Zeno measurements is a random variable, since an unsuccessful projection on $\rho_g$ can happen, in spite of a very small occurrence probability for such phenomenon.}, while $\beta$ denotes the inverse temperature of the thermal reservoir allowing for the erasure of the memory after each measurement.

Since the battery is an open quantum system, its evolution entails energy leakages, which are equal to $\Delta L_{\textrm{evol}} = \int\text{Tr}\left[H_{0}\mathcal{D}[\rho_t]\right]dt$, the integrated energy flow between $\mathcal{B}$ and the environment. In contrast, the projective measurements, assumed as discontinuous operations, are not affected by the environment. This assumption is clearly just an abstraction, which may be extended to non-ideal measurements with an inherent energy consumption~\cite{Guryanova2018}. During the Zeno protection procedure, losses are on average equal to $\langle\Delta L_{\textrm{Zeno}}\rangle = \sum_{k}\int_{t_k}^{t_k + \tau}\text{Tr}\left[H_{0}D(\rho_t)\right]dt$, where $\tau$ is the time interval between two consecutive Zeno measurements. The analytical expressions of $\langle W_{\textrm{stab}}(t_{\rm fin})\rangle$ and $\langle L(t_{\rm fin})\rangle$ can be found in the SM.

The minimal requirement for the battery stabilization is the equality between the average total work $\langle W_{\textrm{stab}}(t_{\rm max})\rangle$ and the accumulated losses until the time instant $t_{\rm max}$ denoted as \emph{break-even time}. An upper bound of the break-even time can be computed just by inverting the relation $\langle W_{\textrm{stab}}(t_{\rm max})\rangle = E_{\rm max}$. However, to make $\mathcal{B}$ a high-performance battery, the collapse probability $P_g$ has to be as close as possible to zero in the shortest time interval, and this is in contrast with the need of low energy consumption. Thus, during the initialization step of the procedure, a \emph{trade-off} between precision and energy cost is unavoidable. Similar results are also observed during the Zeno protection protocol. In this regard, let us consider the average stabilization power $\langle\mathfrak{P}_{\textrm{stab}}\rangle = \langle \dot{W}_{\textrm{stab}}\rangle \approx \langle W_{\textrm{stab}}(\tau)\rangle/\tau$. As formally proven in the SM, $\langle\mathfrak{P}_{\textrm{stab}}(\tau)\rangle$ is approximately equal to
\begin{equation}\label{power_tau}
\langle\mathfrak{P}_{\textrm{stab}}(\tau)\rangle \approx \left[\,{\rm Tr}[H_{0}(\rho_{i}-\rho_{0})]+\overline{m}\,\beta^{-1}H(P(\rho_{\alpha}))\,\right]/\tau\,,
\end{equation}
where the first and second terms of Eq.\,(\ref{power_tau}) denote, respectively, the average cost per cycle to initially bring the battery close to $\rho_{e}$ and the Landauer cost, spent each cycle to reset the (classical) memory register to the energy values of $\mathcal{B}$. We can thus conclude that the longer $\tau$, the smaller the value of the power required to stabilize on average the battery, but the less accurate will be the precision to bring it on $\rho_{e}$.

\begin{figure}
  \includegraphics[scale=0.56]{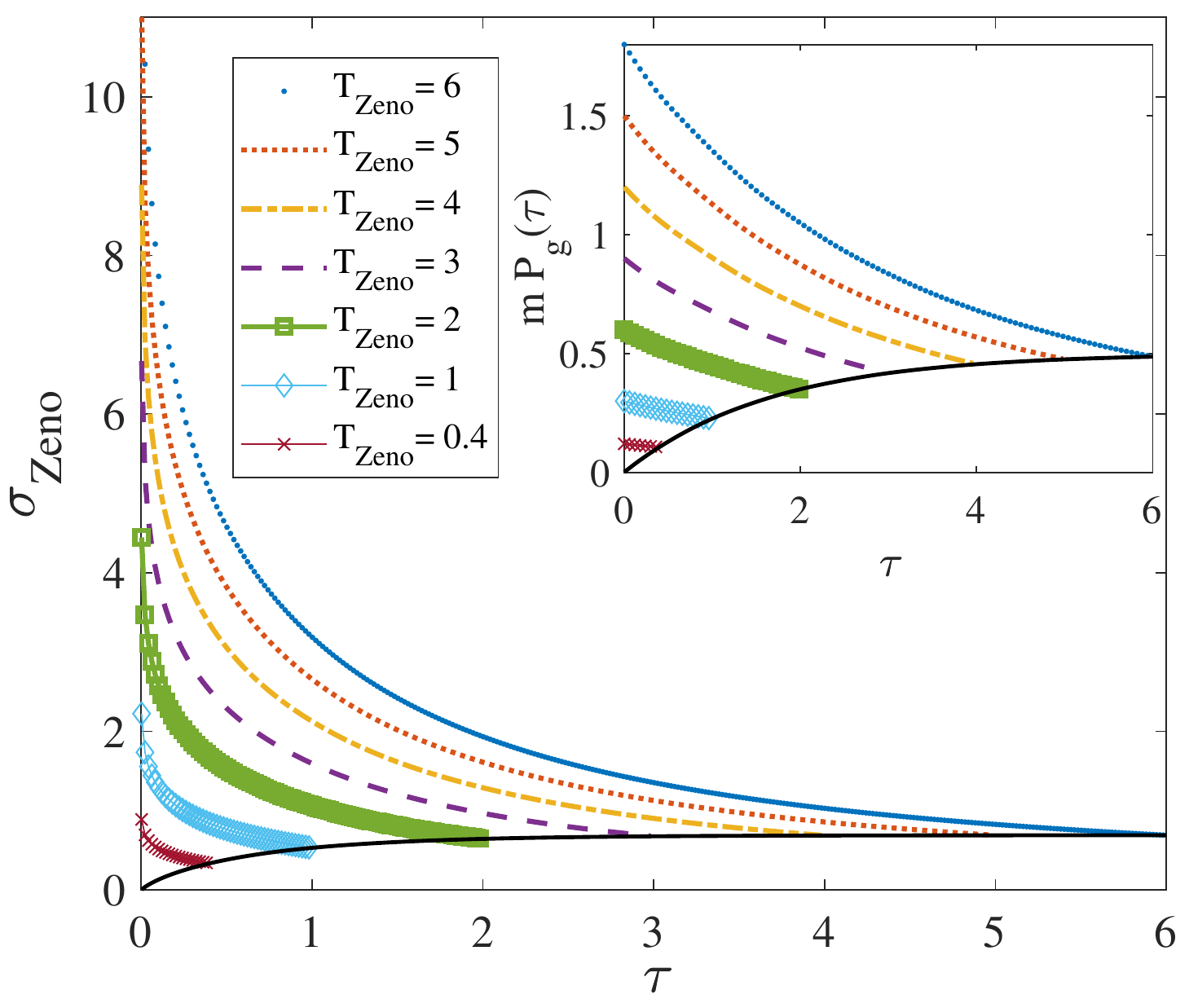}
  \caption{
  Entropic cost of the Zeno protection procedure: $\sigma_{\rm Zeno}$ as a function of the time interval $\tau$ between Zeno measurements. The results have been numerically derived for the same quantum system used in Fig.\,\ref{fig1_main}. Each curve has been obtained by choosing a fixed duration $T_{{\rm Zeno}} \equiv t_{\rm fin} - t_{\rm max}$, among a set of values (see the legend of the figure), and letting vary $\tau$, so that also $\overline{m} \approx T_{{\rm Zeno}}/\tau$ of Zeno measurements changes every time. The integral $\int_{0}^{\tau}\dot{H}(P)dt$ (black line), numerically solved with initial condition $\rho = \rho_e$, has a monotonically increasing behaviour for greater values of $\tau$, thus identifying $\overline{m}$ as the dominant factor. Inset: Amount of not stored energy $\overline{m}P_{g}(\tau)$ (normalized by $E_{\rm max}$) as a function of $\tau$ for $T_{\rm Zeno}=0.4,1,2,3,4,5,6$, in natural units. Here, the black line denotes $P_{g}(\tau)$, and an unavoidable worsening of the battery stabilization is observed when $\tau$ increases.
  }
  \label{fig2_main}
\end{figure}

\paragraph*{Entropic cost in the Zeno regime.--}

Now, let us analyze in more detail the entropic cost of the Zeno protection procedure, based on applying a sequence of projective energy measurements. In this regard, at the level of the battery, the shorter $\tau$ the closer the density operator of $\mathcal{B}$ approaches $\rho_e$. However, such condition does not imply that the global variation of energy during the procedure decreases in the same way. Indeed, the smaller $\tau$, the greater the average number of Zeno measurements, each of them entailing an entropic cost proportional to the irreversible loss of the information content. From a purely dynamical point of view, this corresponds to the cost of purifying the state of the open quantum battery in correspondence to $\rho_e$. As a result, an increasing of the Landauer cost, corresponding to a more frequent memory erasure, is expected. Such behaviour is illustrated by Fig.\,\ref{fig2_main}, in which the entropy production $\sigma_{\rm Zeno} \equiv m\int_{0}^{\tau}\Sigma_{NU}(\rho_t)dt = m\int_{0}^{\tau}\dot{H}(P)dt$ is plotted as a function of $\tau$ by fixing the duration $T_{\rm Zeno}$ of the procedure. In Fig.\,\ref{fig2_main}, the Landauer entropic cost, being proportional to $1/\tau$, diverges as $\tau$ decreases to zero, and the behaviour of $\sigma_{\rm Zeno}$ as a function of $\tau$ is a decaying exponential. Its exponent has the dimensionality of an energy (in natural units); thus, it represents the global energy variation of $\mathcal{B}$ in the Zeno regime. In summary, the value of $\tau$ has to follow a \emph{trade-off} condition: $\tau$ cannot be too small so as to prevent high energy and entropic costs, but neither too large in order to avoid an inadequate value of the stabilization fidelity.

\paragraph*{Conclusion.--}

Thermodynamics and control theory have been combined with the aim to stabilize an open quantum system that acts as a battery.
The proposed method can be seen as a procedure to make the energy (Hamiltonian) basis of the system a decoherence-free-subspace~\cite{LidarReview2003,DFS_Science,Mueller_Zeno_protocols}. This implicitly corresponds to having engineered a super-operator $\mathcal{C}[\rho_t]$, modeling on average the effects of applying projective measurements, so that $\dot{\rho}_{t} = -i[H_{0},\rho_t] + \mathcal{D}[\rho_t] + \mathcal{C}[\rho_t] \approx 0$, with $\rho_t \approx \rho_e$ for any $t$.
As main outlook, one could take into account the possibility that the projective measurements adopted in the stabilization scheme are \emph{non-ideal}, at least according to the definitions recently introduced in~\cite{Guryanova2018}, so as to prevent an unbounded energy cost for their performance. Another promising direction for future research may also lie in the explicit treatment of charging and stabilization fluctuations, as was recently done for Gaussian quantum batteries~\cite{Friis2017}, and the adoption of optimal quantum control theory~\cite{BrifNJP2010} to improve the stabilization procedure.

\begin{acknowledgements}
This work was financially supported by the Fondazione CR Firenze through the project Q-BIOSCAN and PATHOS EU H2020 FET-OPEN grant no.\,828946. F.C.B.\,acknowledges funding from the European Union’s Horizon 2020 research and innovation programme under the Marie Skłodowska‐Curie grant agreement No 801110 and the Austrian Federal Ministry of Education, Science and Research (BMBWF).
\end{acknowledgements}

%%% Bibliography ---------------
\bibliography{stabil_OQB}

%\clearpage

\renewcommand{\thepage}{\roman{page}}
\setcounter{page}{1}
\setcounter{equation}{0}
\setcounter{figure}{0}

        \renewcommand{\theequation}{S\arabic{equation}}%
        \renewcommand{\thefigure}{S\arabic{figure}}%

\begin{widetext}

\section*{Supplemental Material for ``Stabilizing Open Quantum Batteries by Sequential Measurements''}

\subsection*{I. A second-law like inequality for stabilizing an OQB}

In this Supplemental Material (SM), we provide more details about the derivation of the minimum power required to stabilize an OQB affected by decoherence. Our proof follows a similar procedure to the analysis presented in Ref.~\cite{Horowitz2015}.

Let us consider a quantum system dynamics described by the dynamical semi-group $\mathcal{V}(t) \equiv e^{\mathcal{L}t}$ with a (not necessarily unique) fixed point $\pi = \mathcal{V}(t)\pi$ and time-independent $\mathcal{L}$. As shown in~\cite{Breuer2002}, the corresponding entropy production rate $\Sigma(\rho_t)$ related to the battery density operator at time $t$ is \emph{convex} and given by the following relation:
\begin{equation}
 \Sigma(\rho_t) \equiv -{\rm Tr}[\mathcal{L}[\rho_t](\log\rho_t-\log\pi)]\geq 0.\label{eq:er}
\end{equation}
Recasting this general picture to the OQB model discussed in the main text, without applying external control for now, we have $\mathcal{L}[\rho_t] = D[\rho_t]$ and $\pi = \overline{\rho}_{\textrm{dec}}$, where $\overline{\rho}_{\textrm{dec}}$ denotes the (unique) steady-state induced by the presence of battery decoherence alone. Therefore, with our control knob given by a sequence of projective measurements, the entropy production rate $\Sigma(\rho_t)$ of the controlled OQB is greater or equal to the entropy contribution $\Sigma_{D}(\rho_t)$ due to the environment alone. More formally, $\Sigma(\rho_t) = \Sigma_{D}(\rho_t) + \Sigma_{NU}(\rho_t)$, with $\Sigma_{NU}(\rho_t)$ denoting the entropy production rate of the battery given by the non-unitary control transformation. In other words, the total entropy production is lower bounded as
\begin{equation}
\Sigma(\rho_t) \geq -{\rm Tr}[D[\rho_t](\log\rho_t-\log\pi)] \equiv \Sigma_{D}(\rho_t).
\end{equation}
Let us observe that by means of the control procedures the battery is stabilized in the sense that its density operator $\rho_t$ approaches the maximum energy state $\rho_e$, which thus becomes an equilibrium state induced by the control. Moreover, since we are assuming that any operation on $\mathcal{B}$ preserves the trace of its density operator, the total entropy production $\Sigma(\rho_t)$ is non-negative due to the monotonicity of relative entropies under CPTP maps. This means that only energy exchanges are allowed, such that the evolution of the uncontrolled battery can be always described by a CPTP quantum map.

As next step, we quantify the rate of change of the battery total energy $E(\rho_t)$ under stabilizing control by using the first law of thermodynamics, with $\dot{E}(\rho_t)$ given by the relation
\begin{equation}
\dot{E}(\rho_t) = \dot{E}_{D}(\rho_t) + \dot{W}_{\textrm{stab}}(\rho_t),
\end{equation}
where $\dot{E}_{D}(\rho_t)$ is the energy current driven into the battery by the environment, while $\dot{W}_{\textrm{stab}}(\rho_t)$ denotes the power required to charge $\mathcal{B}$ and stabilize it against decoherence. Here, the cost of the sequential measurements is included within the control cost $W_{\textrm{stab}}$. Our goal is to find a lower bound for $\dot{W}_{\textrm{stab}}(\rho_t)$. The energy $E_{D}(\rho_t)$ due to decoherence is given by
\begin{equation}
E_{D}(\rho_t) \equiv {\rm Tr}[\rho_{t}H_{0}] - {\rm Tr}[\rho_{0}H_{0}],
\end{equation}
and the corresponding infinitesimal energy leakage is equal to
\begin{equation}
\dot{E}_{D}(\rho_t) = {\rm Tr}[D[\rho_t]H_0].
\end{equation}
If the battery is a two-level system, the energy current $\dot{E}_{D}(\rho_t)$ can be written as
\begin{equation}
\dot{E}_{D}(\rho_t) = - T_{\overline{\rho}_{\textrm{dec}}}{\rm Tr}[D[\rho_t]\log\overline{\rho}_{\textrm{dec}}] = \dot{E}(\rho_t) -\dot{W}_{\textrm{stab}}(\rho_t),
\end{equation}
where $T_{\overline{\rho}_{\textrm{dec}}}$ is the effective temperature of the battery in correspondence of the steady-state $\overline{\rho}_{\textrm{dec}}$. We point out that $T_{\overline{\rho}_{\textrm{dec}}}$ is more than a parameter introduced for mathematical convenience: it corresponds to the physical temperature of a fictitious quantum system leading to the same decoherence effect as the general dynamics assumed here.

In this way, a lower-bound for $\dot{W}_{\textrm{stab}}(\rho_t)$ can now be derived. We first reconsider Eq.~(\ref{eq:er}) again for the OQB in absence of control:
\begin{equation}
\Sigma_{D}(\rho_t) = -{\rm Tr}[D[\rho_t]\log\rho_t] + {\rm Tr}[D[\rho_t]\log\overline{\rho}_{\textrm{dec}}]\geq 0.
\end{equation}
Since $-{\rm Tr}[D[\rho_t]\log\rho_t] = \dot{S}_{D}(\rho_t)$ is the time-derivative of the von-Neumann entropy for the uncontrolled battery, we have that
\begin{equation}
\dot{W}_{\textrm{stab}}(\rho_t) - \dot{E}(\rho_t) \geq -T_{\overline{\rho}_{\textrm{dec}}}\dot{S}_{D}(\rho_t),
\end{equation}
which leads to the analytical expression of the lower-bound of $\dot{W}_{\textrm{stab}}$:
\begin{equation}\label{lower-bound-app}
\dot{W}_{\textrm{stab}}(\rho_t) \geq \dot{E}(\rho_t) - T_{\overline{\rho}_{\textrm{dec}}}\dot{S}_{D}(\rho_t).
\end{equation}
Notice that Eq.\,(\ref{lower-bound-app}) has to fulfill the second law of thermodynamics. Indeed, by substituting $\dot{E} = \dot{E}_{D} + \dot{W}_{\textrm{stab}}$ from the first law of thermodynamics, the lower-bound (\ref{lower-bound-app}) can be recast in the following inequality:
\begin{equation}\label{inequality_dot_F}
\dot{F}(\rho_t) \equiv \dot{E}_{D}(\rho_t) - T_{\overline{\rho}_{\textrm{dec}}}\dot{S}_{D}(\rho_t) \leq 0,
\end{equation}
where $F(\rho_t)$ stands for the free-energy of the uncontrolled battery. The inequality\,(\ref{inequality_dot_F}) implies that, without controlling the battery, its entropy unavoidably grows due to decoherence, leading thus to a progressive decreasing of the battery free-energy.

Here, it is worth observing that the lower-bound (\ref{lower-bound-app}) is quite conservative, in the sense that the value provided for the minimum control power $\dot{W}_{\textrm{stab}}$ could be overestimated. This is because we have not directly expressed $\dot{W}_{\textrm{stab}}$ as a function of the entropy production rate $\Sigma_{NU}(\rho_t)$ given by controlling $\mathcal{B}$ by means of the proposed non-unitary transformation (sequence of projective measurements). $\Sigma_{NU}(\rho_t)$ is equal to the sum of the entropy production rates associated to each projective measurement, and it can be obtained by evaluating the energy cost in storing and erasing the measurement outcomes in relation to Landauer's principle~\cite{LorenzoPRL2015,MancinoNPJ2018,Abdelkhalek2016}. In this regard, by considering the expression for the entropic contribution $s_t$ of each single measurement result, i.e., $s_t \equiv -\log P$, with $P$ (equal to $P_e$ or $P_g$) denoting the probability that the battery collapses in one of the two energy eigenstates~\cite{Strasberg2018}, we find that
\begin{equation}
\Sigma_{NU}(\rho_t) = \dot{H}(P) = -\dot{P}_{e}\log\left(\frac{P_{e}}{1-P_{e}}\right),
\end{equation}
where $\dot{H}$ denotes the time-derivative of the Shannon entropy $H(P) \equiv - \sum_{k\in\{e,g\}}P_{k}\log P_{k}$, with $P_g = 1 - P_e$. It is worth noting that the entropy production $\Sigma_{NU}$ is zero if and only if $P_g = P_e = 1/2$, i.e., the probabilities that the battery collapses in the maximum or lowest energy state are both equal to $1/2$.

\subsection*{II.\,Energetic balance equation}

In this paragraph we provide more details on the energetic balance equation for an open quantum battery controlled by a sequence of projective measurements. As explained in the main text, the energetic balance equation is evaluated in the limit of fast control, i.e., $V(t)=0$ almost for any $t$. We separately characterize the average control energy $\langle W_{\textrm{stab}}(t_{\rm fin})\rangle$ and the energy leakage $\langle L(t_{\rm fin})\rangle$ within the total time interval $[0,t_{\rm max}]$.

Regarding $\langle W_{\textrm{stab}}(t_{\rm fin})\rangle$, the energy costs to initialize the battery and apply a projective measurement in a single realization of the procedure are respectively equal to
\begin{equation}
\Delta E_{\textrm{evol}} = {\rm Tr}[H_0(\rho_{i}-\rho_0)]
\end{equation}
and
\begin{equation}
\Delta E_{\textrm{meas}} =
\begin{cases}
{\rm Tr}[H_0(\rho_e - \rho_{i})],\,\,\,\text{with probability}\,\,\,P_{e} = {\rm Tr}[\rho_{i}\rho_{e}] \\
{\rm Tr}[H_0(\rho_{i}-\rho_g)],\,\,\,\text{with probability}\,\,\,P_{g} = 1-P_{e}
\end{cases}.
\end{equation}
$\langle\Delta E_{\textrm{meas}}\rangle$ approaches zero on average, as argued in the main text. Instead, the entropic cost for the erasure of the measurement information is equal to $\beta^{-1}H(P)$, where $\beta$ denotes the inverse temperature associated to the thermal reservoir allowing for the resetting of the memory after each measurement. This also means that in the Zeno regime the energy cost of a projection on the energy basis is equal on average to $\beta^{-1}H(P(\rho_{\alpha}))$, such that overall one has that
\begin{equation}
\langle\Delta E_{\textrm{Zeno}}\rangle = \overline{m}\,\beta^{-1}H(P(\rho_{\alpha})),
\end{equation}
where $\overline{m}$ denotes the average number of Zeno measurements and $\rho_{\alpha}$ is the average state of the battery immediately before each measurement. Note that the average measurement cost $P_{e}(\rho_{\alpha}){\rm Tr}[H_0(\rho_e - \rho_{\alpha})] + P_{g}(\rho_{\alpha}){\rm Tr}[H_0(\rho_g - \rho_{\alpha})]$ during the Zeno protection procedure is vanishing. As a result, since the stabilization procedure is repeated with probability $P_e$ until the battery is charged, i.e., $\rho_t$ reaches $\rho_e$, the average total work needed to keep the energy storage in the battery until $t=t_{\rm fin}$ is given by the following relation:
\begin{equation}\label{total_work}
\langle W_{\textrm{stab}}(t_{\rm fin})\rangle \approx \left(1+\sum_{k=1}^{\overline{N}}P_{g}^{k}\right)\Delta E_{\textrm{evol}} + \langle\Delta E_{\textrm{Zeno}}\rangle,
\end{equation}
where $\overline{N}$ is the average number of times the stabilization procedure is repeated with probability $P_g$.

Then, let us derive the average total energy leakages $\langle\Delta L(t_{\rm fin})\rangle$. By considering each projective measurement as a discontinuous operation, the measurement process is not affected by the environment. Thus, only the energy leakages during the dynamics of the battery and the Zeno protection stage have to be considered. In this regard, the former is given by
\begin{equation}
\Delta L_{\textrm{evol}} = \int\text{Tr}\left[H_{0}\mathcal{D}[\rho_t]\right]dt,
\end{equation}
while the latter on average is globally equal to
\begin{equation}
\langle\Delta L_{\textrm{Zeno}}\rangle = \sum_{k=1}^{\overline{m}}\int_{t_k}^{t_k + \tau}\text{Tr}\left[H_{0}D(\rho_t)\right]dt,
\end{equation}
where $\tau$ is the time interval between two consecutive Zeno measurements. In conclusion, this implies that the average total energy leakage $\langle\Delta L(t_{\rm fin})\rangle$ at the final time instant $t_{\rm fin}$ is
\begin{equation}\label{total_losses}
\langle\Delta L(t_{\rm fin})\rangle \approx \left(1+\sum_{k=1}^{\overline{N}}P_{g}^{k}\right)\Delta L_{\textrm{evol}} + \langle\Delta L_{\textrm{Zeno}}\rangle,
\end{equation}
with $\Delta L_{\textrm{evol}}$ counted $\overline{N}$ times until $\rho_t = \rho_e$.

As final remark, it is worth observing that for a vanishing value of $P_g$ we can perform a first-order expansion of both $\langle W_{\textrm{stab}}(t_{\rm fin})\rangle$ and $\langle\Delta L(t_{\rm fin})\rangle$ as a function of $P_{g}$, so that $\langle W_{\textrm{stab}}(t_{\rm fin})\rangle \approx (1+P_g)\Delta E_{\textrm{evol}} + \langle\Delta E_{\textrm{Zeno}}\rangle$ and $\langle\Delta L(t_{\rm max})\rangle \approx \left(1+P_{g}\right)\Delta L_{\textrm{evol}} + \langle\Delta L_{\textrm{Zeno}}\rangle$. Therefore, if we also reasonably assume that in the Zeno regime the sum of the energy losses is on average almost equal to the energy required to charge the quantum system (i.e., $\langle\Delta E_{\textrm{Zeno}}\rangle \approx \langle\Delta L_{\textrm{Zeno}}\rangle$), the relative excess stabilization cost at $t = t_{\rm fin}$ is given by the following relation:
\begin{equation}
\xi_{\textrm{stab}}(t_{\rm fin}) = \frac{1+P_g}{E_{\textrm{max}}}\left|\Delta E_{\textrm{evol}} - \Delta L_{\textrm{evol}}\right|,
\end{equation}
with the result that $\xi_{\textrm{stab}}(t_{\rm fin}) = 0$ if the energy cost to drive the battery up to the state $\rho_{i}$ perfectly equals the decoherence losses during the battery evolution (i.e., if the losses are just compensated by the control action).

\subsection*{III.\,Details about the numerical implementation}

The results of Fig.\,\ref{fig1_main} in the main text have been obtained by considering as quantum battery the following two-level system, with internal Hamiltonian
\begin{equation}
H_0 = \Omega\sigma_x + \omega\sigma_z,
\end{equation}
where $\omega = 1$, $\Omega = 3$ (in units such that $\hbar=1$), and $\sigma_x$, $\sigma_z$ Pauli matrices. Thus, in the basis of $\sigma_z$, given for convention by the eigenstates $|0\rangle \equiv [0,1]^{T}$ and $|1\rangle \equiv [1,0]^{T}$ (the superscript $(\cdot)^{T}$ denotes the transposition symbol), the corresponding maximum and minimum energy states are respectively equal to
\begin{equation}
\rho_e \approx \begin{pmatrix}
0.658  &  0.474 \\
0.474  &  0.342
\end{pmatrix}\,\,\,\text{and}\,\,\,\rho_g \equiv \mathbb{I}_{\mathcal{B}} - \rho_e \approx \begin{pmatrix}
0.342   &  -0.474 \\
-0.474  &  0.658
\end{pmatrix}.
\end{equation}
Here, we have reasonably chosen as input density operator $\rho_0$ the minimum energy state $\rho_g$. Moreover, in order to fulfill the \emph{fast control} condition, we have assumed to use a time-dependent term $V(t)$ in the driving Hamiltonian only to slightly bring out-of-equilibrium the battery from $\rho_g$ to $\equiv |0\rangle\!\langle 0|$. Given the internal time-independent Hamiltonian $H_0$, this operation is achieved by taking $V(t) = e^{-i\phi\sigma_{y}}$ (rotation around the y-axis), with
\begin{equation}
\phi \equiv \arctan\left(\frac{\rho_{g}^{(11)}}{\rho_{g}^{(21)}}\right).
\end{equation}
Then, to transfer the battery population from $|0\rangle\!\langle 0|$ to the initialization state $\rho_i$, only the dynamical evolution governed by $H_0$ has been exploited. Accordingly, under this assumption, we need to determine the optimal value of $t$ (i.e., $t^{\ast}$) in correspondence of which it is worth performing the first energy projective measurement of the protocol.

The dynamical evolution of the system is given by the Markovian master equation $\dot{\rho}_t = -i[H_0,\rho_t]+D[\rho_t]$ ($\hbar = 1$), where the super-operator $D[\rho_t]$ modeling decoherence within the battery dynamics has been chosen equal to
\begin{equation}
D[\rho_t] = \gamma\left(-\{\mathcal{N},\rho_t\}+2\mathcal{N}\rho_{t}\,\mathcal{N}\right),
\end{equation}
i.e., as an operator inducing pure-dephasing, with $\{\cdot,\cdot\}$ Poisson bracket, $\mathcal{N} \equiv |1\rangle\!\langle 1|$ and $\gamma = 2/3$. Here, it is worth noting that also pure-dephasing master equations, despite they are energy preserving, involve dynamical behaviours worthy of being studied, since stabilizing an OQB implicitly implies the protection (in our case) of coherence in the battery energy basis. Moreover, the motivation under the choice of $\gamma = 2/3$ will be clear below. In Fig.~\ref{fig1_app} we show the behaviour over time of both the populations and coherence of the two-level system for a whole duration of the dynamics taken equal to $10$ (always in natural units) by starting from the state $|0\rangle\!\langle 0|$.
\begin{figure}[h!]
\includegraphics[width=0.475\linewidth]{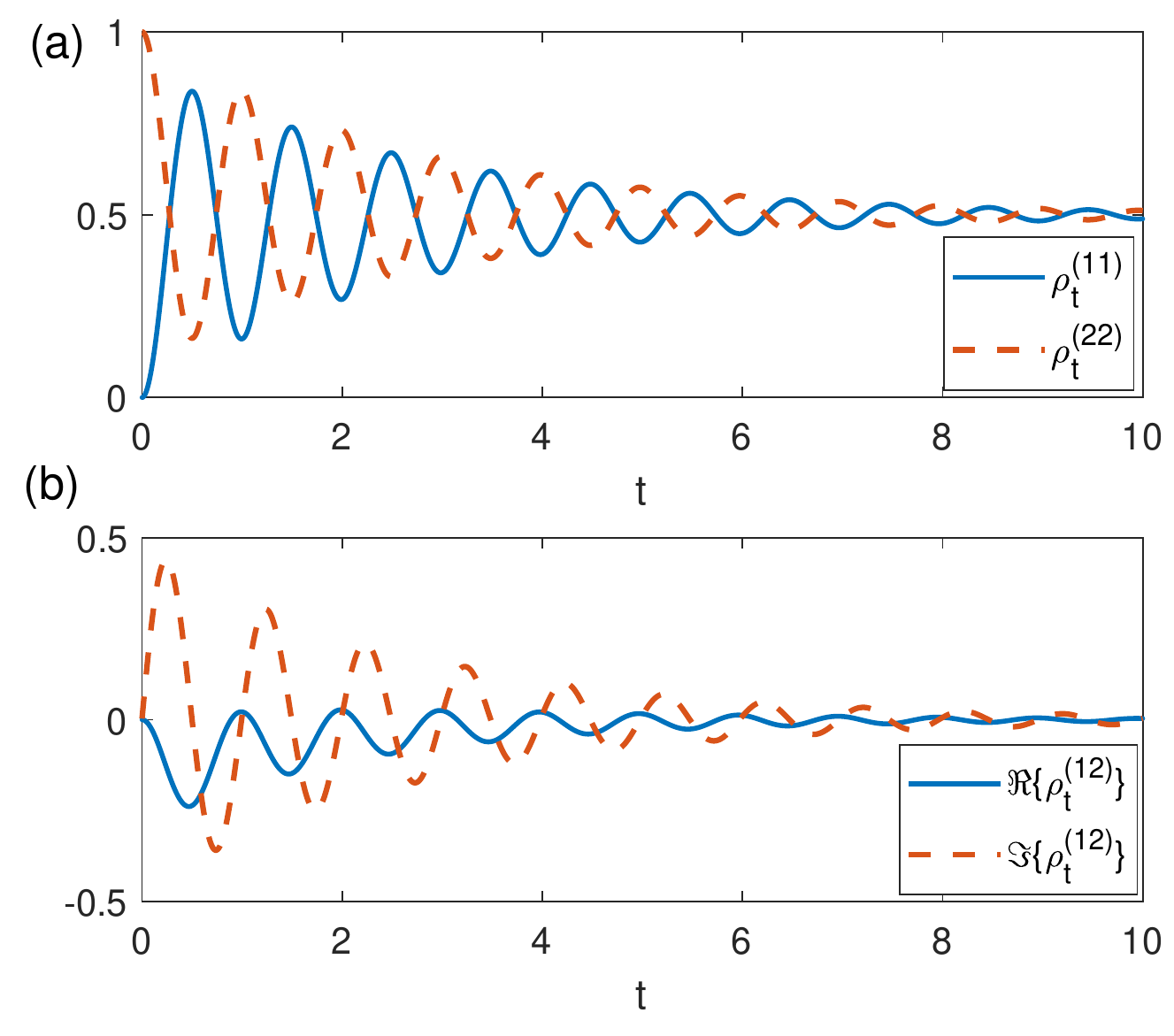}
\caption{(a)\,Behaviour over time of the battery density matrix elements $\rho^{(11)}_t$ and $\rho^{(22)}_t = 1 - \rho^{(11)}_{t}$. (b)\,Behaviour over time of the real and imaginary part of the battery density matrix element $\rho^{(12)}_t$.}
\label{fig1_app}
\end{figure}
Instead, in Fig.\,\ref{fig2_app} we plot the trace distance of $\rho_t$ w.r.t.\,$\rho_e$, as well as the probability $P_e(t) \equiv {\rm Tr}[\rho_t\rho_e]$. We can observe that the trace distance $T(\rho_t,\rho_e)$ is always greater or equal to $1/2$, so that $P_g \geq 1/2$.
\begin{figure}[h!]
\includegraphics[width=0.475\linewidth]{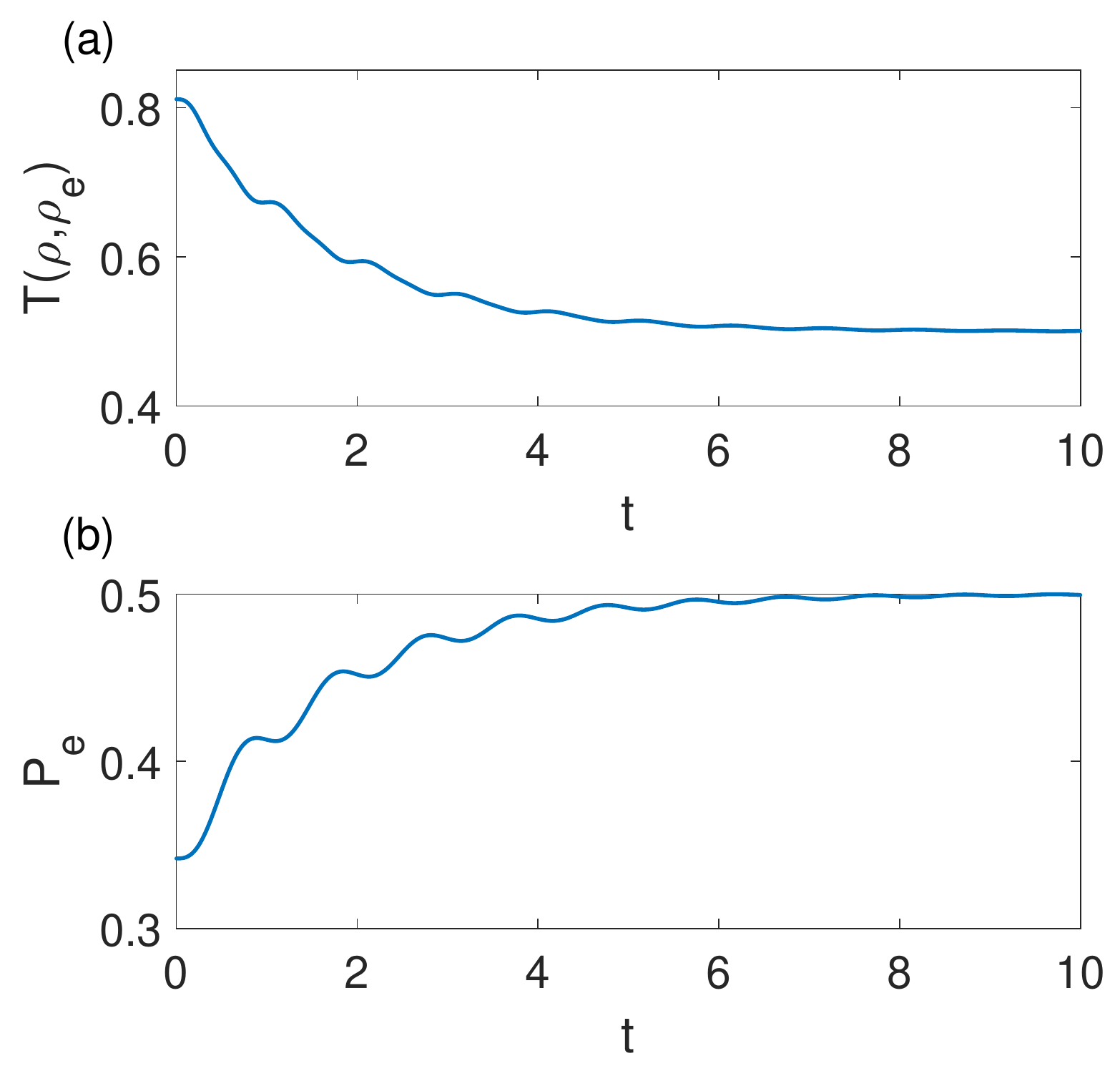}
\caption{(a)\,Behaviour over time of the trace distance $T(\rho_t,\rho_e)$. (b) Behaviour over time of the probability $P_e$.}
\label{fig2_app}
\end{figure}
In this regard, in the numerical simulations we have verified that also in this case $\langle\Delta E_{\textrm{meas}}(t)\rangle \equiv P_g{\rm Tr}\left[H_0(\rho_{t}-\rho_g)\right] + P_e\left[H_0(\rho_e - \rho_{t})\right] \approx 0$ for each $t$ within all the evolution of the system, and thus also for $t^{\ast}$.

The results in Figs.\,\ref{fig1_app},\,\ref{fig2_app} could induce the experimenter to take as $t^{\ast}$ a sufficiently long time interval so as to minimize the trace distance $T(\rho_t,\rho_e)$ and at the same time maximize the probability $P_e$. However, being the stabilization scheme a probabilistic procedure, this choice could bring the main disadvantage to wait for a long time interval and then observe the battery staying for the most of the time not on the maximum energy state but in correspondence of $\rho_g$, and thus leading to a very low stabilization fidelity $\mathcal{F}$. To make a better choice of $t^{\ast}$, it is worth analyzing the reason why $P_e$ is always $\leq 1/2$. We find that, in order to achieve the maximum energy state $\rho_e$, we need to stabilize both populations and coherence of the battery. But with the chosen internal Hamiltonian $H_0$ the stabilization of populations and coherence cannot be reached at the same time. In this regard, there are three possibilities: (a) minimize \emph{only} the difference between the modulus squared of coherence terms of $\rho_t$ and $\rho_e$; (b) minimize \emph{only} the difference between the modulus squared of the diagonal terms of $\rho_t$ and $\rho_e$; (c) find a \emph{trade-off} between (a) and (b) by ensuring that the value of $t^{\ast}$ is not too large and at the same time $\langle W_{\textrm{stab}}\rangle$ is as small as possible w.r.t.\,the average energy leakages $\langle\Delta L\rangle$. For the specific implementation of Fig.~\ref{fig1_main} we have chosen the solution (c) corresponding to a value of $t^{\ast}$ equal to $0.33$ (in natural units). Notice that, being $P_e \leq 1/2$ by starting from the state $|0\rangle\!\langle 0|$, we can at most minimize the average total control work and get a very high fidelity $\mathcal{F}$, but without achieving the best possible energetic efficiency. We have deliberately chosen this example in order to show that the proposed stabilization scheme, based on sequential quantum measurements, turns out to be extremely efficient from an energetic point of view only if the probability to collapse onto $\rho_e$ after each quantum measurement is sufficiently high, ideally close to $1$. If not, a greater energy cost (if compared with $E_{\textrm{max}}$) is required, so as to bring the system into the maximum energy state and at the same time compensate the presence of the external environment leading to decoherence.

Finally, as it can be observed by Fig.~\ref{fig1_main}, the probability that the state of the battery collapses in the minimum energy state $\rho_g$ while the procedure of Zeno protection is turned on is very low (smaller than 1\%). This is due to our choice to take the time interval $\tau$ between Zeno measurements equal to $0.0662$ (in natural units), 5 times smaller than $t^{\ast}$. However, there does exist the possibility that the Zeno protection procedure would fail; in such a case we simply re-initialize the system and the stabilization scheme is repeated from the beginning. In this regard, it is worth noting that for larger values of $\tau$ the fidelity $\mathcal{F}$ decreases and consequently $\langle W_{\textrm{stab}}\rangle$ unavoidably increases, since for a fixed value of $T_{{\rm Zeno}}$ the stabilization scheme needs be re-initialized a greater number of times.

\end{widetext}

\end{document}